\documentclass[12pt,a4paper]{article} \usepackage[latin1] {inputenc}
\usepackage{amsmath} \usepackage{amsfonts} \usepackage{amssymb} 
\usepackage[dvips]{graphicx,color}
\begin{document}
\title{Fermion propagator for $QED_3$ in the IR domain}
\author{C.~D.~Fosco and A.~L{\'o}pez~\footnote{On leave from Centro
    At\'omico Bariloche, 8400 S.\ C.\ de Bariloche, Argentina.}
  \\
  {\normalsize\it Department of Physics, Theoretical Physics}\\
  {\normalsize\it University of Oxford}\\
  {\normalsize\it Oxford OX1 3NP, United kingdom.} }  \date{\today}
\maketitle
\begin{abstract}
\noindent We evaluate the fermion propagator in parity-conserving
$QED_3$ with $N$ flavours, in the context of an IR domain
approximation. This provides results which are non-perturbative in the
loopwise expansion sense. We include fermion-loop effects, and show
that they are relevant to the chiral symmetry breaking phenomenon,
that can be understood in this context.
\end{abstract}

Parity-conserving $QED$ in $2+1$ dimensions, in the context of the
large-$N$ approximation, is an effective theory description for many
interesting models of condensed matter theory~\cite{qed3}. Moreover,
it provides a simple yet non-trivial model where to study chiral
symmetry breaking. Unlike in $1+1$ dimensions, there is a truly
dynamical gauge field; besides, the model shares some important
properties with its $3+1$ dimensional counterpart. For example, the
dimensionful coupling constant for the $2+1$ dimensional model plays
an analogous role to the renormalization scale of the $3+1$
dimensional theory.  It is, however, a much simpler framework where to
study chiral symmetry breaking, and it allows for the use of
controlled approximations.  Many detailed studies have been carried
out using the truncated Schwinger-Dyson equations for the gap function
$\Sigma$, both at zero~\cite{Pisarski:dj}-\cite{Nash:1989xx} and
finite~\cite{Dorey:1991kp}-\cite{Lee:1997xf} temperature.  Results
(which are consistent with the continuum calculations) have also been
obtained by using numerical lattice
calculations~\cite{Dagotto:1988id}-\cite{Hands:2002qt}.

It is well known that $QED_3$ has a non trivial IR structure, where
the {\em a priori\/} severe IR divergences can be consistently treated
thanks to the model superrenormalizability~\cite{j-t}. Indeed, the
leading and first sub-leading logarithms have been summed to all
orders for the fermion self-energy~\cite{temp}.

The relevant degrees of freedom driving the chiral phase transition
are the low-momentum ones, with an effective dynamical cutoff set up
by the (square of the) coupling constant.  It then makes sense to
attempt approximations based on that property.  In this note, we
calculate the full fermion propagator in parity-conserving large-$N$
$QED_3$, in the context of the IR domain technique presented
in~\cite{Karanikas:1995zi}, and comment on the relevance of this
result for the chiral phase transition.

The model may be defined in terms of an Euclidean generating
functional ${\mathcal Z}(j,{\bar\eta},\eta)$, which in the
path-integral representation is given by
\begin{equation}\label{eq:defz}
{\mathcal Z}(j,{\bar \eta},\eta )=\int [{\mathcal D}A ] {\mathcal
D}{\bar\psi}{\mathcal D}\psi \, \exp \left[- S({\bar\psi},\psi,A) + \int
d^3x( j_\mu A_\mu + {\bar\eta}\psi + {\bar\psi} \eta )\right]\;,
\end{equation}
where the Euclidean action $S$ is,
\begin{equation}\label{eq:defs}
S({\bar\psi},\psi,A)\;=\;S_f({\bar\psi},\psi,A) \,+\, S_g (A)
\end{equation}
with $S_f$ and $S_g$ denoting the fermion and gauge field actions,
respectively. For $S_f$, we assume a Dirac-like form:
\begin{equation}\label{eq:defsf}
S_f({\bar\psi},\psi,A)\;=\; \int d^3x \sum_{a=1}^N \; {\bar\psi}_a
( \not\!\partial + i e_N \not\!\! A ) \psi_a 
\end{equation}
while the gauge field has a standard Maxwell action:
\begin{equation}\label{eq:defsg}
S_g (A) \;=\;\int d^3x \, \frac{1}{4} \, F_{\mu\nu} F_{\mu\nu} \;,
\;\;\;\; F_{\mu\nu}=\partial_\mu A_\nu - \partial_\nu A_\mu \;.
\end{equation}
Here~\mbox{$e_N = e N^{-1/2}$}, where $e$ is a coupling constant (with
the dimensions of the square root of a mass). It is the only
dimensionful parameter in the model. The $N^{-1/2}$ factor is
introduced, as usual in the large-$N$ expansion, in order to have a
finite limit for the leading term in the expansion, when $N \to
\infty$.

The Dirac $\gamma$-matrices are in a (reducible) $4\times 4$
representation of the Dirac algebra:
\begin{equation}
\gamma_0 \;=\; \left( \begin{array}{cc}
                         \sigma_3 & 0 \\
                          0 & - \sigma_3
                         \end{array} \right)
\;\;,\;\;
\gamma_1 \;=\; \left( \begin{array}{cc}
                         \sigma_1 & 0 \\
                          0 & - \sigma_1
                         \end{array} \right)
\;\;,\;\;
\gamma_2 \;=\; \left( \begin{array}{cc}
                         \sigma_2 & 0 \\
                          0 & - \sigma_2
                         \end{array} \right)
\end{equation}
where $\sigma_j$, $j=1,2,3$ are the standard Pauli matrices. In this
representation, the model has an extra `chiral' symmetry, generated by
a `$\gamma_5$' matrix
\begin{equation}\label{eq:defgamma5}
\gamma_5 \;=\; \left( \begin{array}{cc}
                   0     &    {\mathbf 1}_{2\times 2} \\
                 {\mathbf 1}_{2\times 2} &      0
                      \end{array} \right) \;,
\end{equation} 
which together with other generators complete the Lie algebra of an
internal $U(2N)$ symmetry group.  This symmetry is known to be
dynamically broken\cite{Pisarski:dj}, and a non-vanishing mass for the
fermions is generated as a consequence. The $U(2N)$ symmetry breaks
down to $U(N) \times U(N)$.

Besides, a fermion mass {\em does not\/} break the parity symmetry of
the massless model.  This means that parity symmetry is not anomalous,
since there are regularizations that indeed preserve that symmetry,
like the Pauli-Villars method, for example.  This guarantees that
there is no radiatively induced Chern-Simons term, and that it is
perfectly consistent to use a parity-conserving action, since parity
is not broken by renormalization effects.

The symbol $[{\mathcal D}A_\mu]$ in (\ref{eq:defz}) denotes the gauge
field functional integration measure, plus any gauge-fixing artifacts.
An effect of the breaking of chiral symmetry is given by the
generation of a non-vanishing mass gap $\Sigma$ for the fermionic
field. This may in turn be observed by evaluating the full fermion
propagator $G$ in the theory defined by (\ref{eq:defz})
\begin{equation}\label{eq:defg}
G(x-y) \;=\; \langle \psi (x) {\bar\psi}(y) \rangle  \;,
\end{equation}
which, (by integrating out the fermion fields) may be written as
\begin{equation}\label{eq:defg1}
G(x-y) \;=\; \frac{1}{{\mathcal N}} \int [{\mathcal D}A] \,
\langle x| {\not \!\! D}^{-1} |y \rangle \, (\det \not \!\! D)^N \,
\exp [ - S_g(A) ] \;,
\end{equation}
where
\begin{equation}\label{eq:defg2}
{\mathcal N}\;=\;  \int [{\mathcal D}A] \, (\det \not \!\! D)^N \,
\exp [ - S_g(A) ] \;,
\end{equation}
and we have adopted Schwinger's notation \mbox{$\langle x|K|y
  \rangle$} for the coordinate space matrix elements (or kernel) of an
operator $K$; in the case above the fermion propagator in an external
gauge field.

In the large-$N$ limit, the leading contribution from the fermionic
determinant comes from its quadratic part, namely,
\begin{equation}\label{eq:fdet}
(\det \not \!\! D)^N \,=\, \exp ( N {\rm Tr} \ln  \not \!\! D )
\,=\, \exp[- W^{(2)}(A) + {\mathcal O}(N^{-1}) ]
\end{equation}
where
\begin{equation}
 W^{(2)}(A) \,=\, - \frac{e^2}{2} \, {\rm Tr} \left[ {\not \! \partial}^{-1} \not\!\!A
 {\not \! \partial}^{-1} \not \!\!A \right] \;. 
\end{equation}
The exact form of the quadratic functional $W^{(2)}(A)$ is well known;
it may be written as follows:
\begin{equation}
W^{(2)}(A) \;=\; \frac{e^2}{32} \, \int d^3x \, F_{\mu\nu}
\frac{1}{\sqrt{-\partial^2}} F_{\mu\nu}
\end{equation}
The operator $\frac{1}{\sqrt{-\partial^2}}$, which has a nonlocal
kernel in coordinate space, may be thought of as being defined by its
Fourier representation:
\begin{equation}
\langle x|\frac{1}{\sqrt{-\partial^2}}|y\rangle\;=\; 
\int \frac{d^3k}{(2\pi)^3} e^{i k \cdot (x-y)} \, \frac{1}{|k|} \;,
\end{equation}
where it is obviously local.

Thus, in the large-$N$ limit, the gauge field average of
(\ref{eq:defg1}) has a purely Gaussian weight. The evaluation of the
gauge field average may be further simplified by using a convenient
gauge-fixing term:
\begin{equation}
[{\mathcal D}A]\;=\; {\mathcal D}A \, \exp[ - \frac{1}{2 \xi} 
\int d^3x \; \partial\cdot A \,( 1 + \frac{e^2}{8
  \sqrt{-\partial^2}}) \, \partial \cdot A ] \;,
\end{equation}
where $\xi$ is a real constant. Then
\begin{equation}\label{eq:g1}
G(x-y) \;=\; \frac{1}{{\mathcal N}} \int {\mathcal D}A \,
\langle x| {\not \!\! D}^{-1} |y \rangle \, \exp [ - S_G(A) ] \;,
\end{equation}
where the quadratic functional $S_G(A)$ may be written explicitly as
\begin{equation}
S_G(A) \;=\; \frac{1}{2} \int d^3x \, A_\mu M_{\mu\nu} A_\nu 
\end{equation}
with
\begin{equation}
M_{\mu\nu} \;=\; M(\sqrt{-\partial^2}) \; \left[ \delta_{\mu\nu} +
  (\xi^{-1}-1) 
\frac{\partial_\mu\partial_\nu}{\partial^2} \right]
\end{equation}
and
\begin{equation}
 M(\sqrt{-\partial^2})\;=\; \sqrt{-\partial^2} \; ( \sqrt{-\partial^2}
+ \frac{e^2}{8}) \;.
\end{equation}

In order to evaluate $G$, in particular its self-energy part, we
follow the approach of~\cite{Karanikas:1995zi}, although with a
difference that turns out to be crucial in our case, namely, keeping
the fermion determinant contribution to the effective gauge field
action. We remark that it is totally justified to ignore the
determinant in~\cite{Karanikas:1995zi}, since the fermions are assumed
to have a non-vanishing renormalized mass, and that mass sets up a
scale where to make the relevant IR approximations. In our case,
however, the only scale appears in the one-loop contribution.
Moreover, there will also be a non-vanishing fermion mass, but here it
will appear in a self-consistent way. This self-consistency depends
upon the inclusion of the full one-loop gauge field propagator.
Indeed, this is known to be important in any study of chiral symmetry
breaking in $QED_3$~\cite{Pisarski:dj}, and this calculation is no
exception, as it will become clear in what follows.

Following~\cite{Karanikas:1995zi}, we represent the fermionic
propagator in an external field in terms of a path integral over
trajectories, as follows
$$
\langle x| {\not \!\! D}^{-1}|y \rangle \;=\; \int_{0^+}^\infty dT
\, \int {\mathcal D}x {\mathcal D}p \; \exp[i \int_0^T d\tau p \cdot
{\dot x}] \;\; {\mathcal P} \exp[-i \int_0^T d\tau {\not\!p}(\tau) ]
$$
\begin{equation}\label{eq:aux}
\times \exp[- i e_N \int_{0}^T d\tau {\dot x} \cdot A(x(\tau))] \;, 
\end{equation}
where ${\mathcal P}$ denotes path ordering, and the functional
integral over $x_\mu$ is taken over paths with the boundary condition
$x_\mu(0) = y_\mu$ and $x_\mu(T) = y_\mu$. There is no boundary
condition for the values of $p_\mu(\tau)$ at $0$ and $T$.  It is clear
that the functional average over $A$ will only affect the last factor
in (\ref{eq:aux}). Indeed,
$$
G(x-y) \;=\;\int_{0^+}^\infty dT \, \int {\mathcal D}x {\mathcal
  D}p \; \exp[ i \int_0^T d\tau \, p(\tau) \cdot x(\tau) ]
$$
\begin{equation}\label{eq:g2}
\times \; {\mathcal P} \exp[ -i \int_0^T d\tau \, {\not \! p}(\tau) ] 
\;\; \exp( -{\mathcal F}) \;,
\end{equation}
where ${\mathcal F}$ is a functional of the trajectories $x(\tau)$ and
a function of $T$, and is defined by the average:
\begin{equation}\label{eq:deff}
 \exp\left\{-{\mathcal F}\right\} \;=\; \left\langle 
\exp [- i e_N \int_{0}^T d\tau {\dot x} \cdot A(x(\tau)) ] \right\rangle \;,
\end{equation}
with
\begin{equation}
\langle \ldots \rangle \;=\; \frac{\int {\mathcal D}A \,
\ldots \, \exp [ - S_G(A) ]}{\int {\mathcal D}A \, \exp [ - S_G(A) ]}
\;.
\end{equation}
Being $S_G$ quadratic, the integral over $A$ can of course be
performed exactly. The result for ${\mathcal F}$ may then be written
as follows
\begin{equation}\label{eq:eff1}
{\mathcal F} \;=\; \frac{e_N^2}{2} \int d^3u \int d^3v \;
j_\mu(u) \langle u |M^{-1}_{\mu\nu}|v\rangle j_\nu(v) 
\end{equation}
where we have introduced the current $j_\mu$ associated to the
particle's worldline:
\begin{equation}
j_\mu (u) \;=\; \int_0^T d\tau \, {\dot x}_\mu(\tau) \, \delta[u -
x(\tau)] \;,
\end{equation}
with ${\dot x}_\mu \equiv \frac{dx}{d\tau}$. Taking into account the
fact that the worldline begins at $y$ and ends at $x$, we see that
\begin{equation}
\partial_\mu j_\mu (u) \;=\; \delta (u - y) - \delta (u - x) \;,
\end{equation}
and because of this, we may evaluate the contribution of ${\mathcal
  F}$ due to the longitudinal part of $(M^{-1})_{\mu\nu}$ exactly.
Namely, the longitudinal part of the propagator, which is where all
the gauge-dependence is contained, will contribute to a term, which we
denote ${\mathcal F}_l$, in the following expression for ${\mathcal
  F}$:
\begin{equation}\label{eq:aux1}
{\mathcal F} \;=\; {\mathcal F}_t\;+\; {\mathcal F}_l
\end{equation}
where
\begin{equation}\label{eq:eff2}
 {\mathcal F}_t\;=\; e_N^2 \int_0^T d\tau \int_\tau^T d\tau' \; 
{\dot x}(\tau) \cdot {\dot x}(\tau') \; 
\langle x(\tau) | M^{-1} | x(\tau') \rangle
\end{equation}
and
\begin{equation}\label{eq:g3}
 {\mathcal F}_l\;=\;e_N^2 \,(\xi - 1) \left[ 
\langle x| {\frac{M^{-1}(\sqrt{-\partial^2}) }{-\partial^2}} |x\rangle  
-  \langle x| {\frac{M^{-1}(\sqrt{-\partial^2}) }{-\partial^2}} |y\rangle \right]\;.
\end{equation}
Each separate term in (\ref{eq:g3}), when written in momentum space,
is divergent. However, their sum may be evaluated by combining the two
integrands under a common momentum integral, what yields a convergent
result. The reason is, of course, that the first term in (\ref{eq:g3})
exactly kills the logarithmic divergent of the second.

The momentum integration can then be performed, obtaining:
\begin{equation}
{\mathcal F}_l[r]\;=\;  \frac{4}{\pi^2 N } \,(\xi -1)
f_l(r) \;,
\end{equation}
where
$$
f_l(r)\;=\; \left\{ \gamma - 1 + \ln (\frac{8 r}{e^2}) +
  \frac{\pi}{2} [ 1 - \cos(\frac{8 r}{e^2})] \right.
$$
\begin{equation}
\left. +\; \frac{8}{e^2 r} [ \cos(\frac{e^2r}{8}) 
\rm{si}(\frac{e^2r}{8})-  \sin(\frac{e^2r}{8}) \rm{ci}(\frac{e^2r}{8})]
\right\} \;,
\end{equation}
where $\gamma$ is the Euler's constant, $r \equiv |x-y|$, and
$\rm{si}$ ($\rm{ci}$) denotes the integral sine (cosine) function.

Since ${\mathcal F}_l$ does not depend on the trajectories $x(\tau)$,
it is clear that the only contribution relevant to the fermion mass
will come from ${\mathcal F}_t$ in (\ref{eq:eff2}). To calculate
${\mathcal F}_t$, we follow the procedure described
in~\cite{Karanikas:1995zi}.  After the approximations are implemented,
and a `ribbon' regularization with a small-distance cutoff $b$ is
introduced~\cite{Karanikas:1995zi}, we end up with
\begin{equation}\label{eq:ft1}
{\mathcal F}_t\;=\; \frac{e^2}{N} \int_0^T d\tau \int_\tau^T d\tau' \; 
M^{-1}(\sqrt{|\tau - \tau'|^2 + b^2})
\end{equation}
where $M^{-1}(x)\equiv\langle x|M^{-1}|0\rangle$.  The kernel
$M(\sqrt{-\partial^2})$ in coordinate space can be computed, yielding
\begin{equation}\label{eq:mx}
M^{-1}(x) \;=\; \frac{1}{2 \pi^2 r} \, 
\left[ 
\frac{\pi}{2} \cos(\frac{e^2r}{8}) \,+\, 
\rm{ci}(\frac{e^2r}{8})
\sin(\frac{e^2r}{8}) \,-\, \rm{si}(\frac{e^2r}{8}) 
\cos(\frac{e^2r}{8}) \right]
\end{equation}
where $r\equiv |x|$. It is evident that, if we use the exact
expression for $M^{-1}(x)$ in (\ref{eq:ft1}), we cannot evaluate the
integral over $\tau$ and $\tau'$ analytically, something that we do
need in order to obtain the fermion self-energy.  One can easily check
that the small-$r$ behaviour of (\ref{eq:mx}) corresponds to the case
of a pure Maxwell gauge field action, namely,
\begin{equation}\label{eq:small}
M^{-1}(x) \;\sim \; \frac{1}{4\pi r} \;,\;\;\; r \sim 0\;, 
\end{equation}
while it yields the pure one-loop action dependence for large-$r$:
\begin{equation}\label{eq:big}
M^{-1}(x) \;\sim \; \frac{4}{\pi^2 e^2 r^2} \;,\;\;\; r >> e^{-2}\;. 
\end{equation}
It is possible to use a simple analytic approximation to
(\ref{eq:mx}), which however captures both (\ref{eq:small}) and
(\ref{eq:big}) exactly and does not differ significantly from the
exact function in between. We use an approximant to $M^{-1}(x)$, which
we denote by ${\mathcal M}^{-1}(x)$, and is given by the expression:
\begin{equation}\label{mapp}
{\mathcal M}^{-1}(x) \;=\; \frac{1}{2 \pi^2 r ( \frac{e^2}{8} r +
    \frac{2}{\pi} ) }\;.
\end{equation}
Inserting this expression in place of the exact one, $M^{-1}$ in
(\ref{eq:ft1}), one can evaluate exactly the integral over $\tau$ and
$\tau'$. Of course, not all the dependence in $b$ is interesting,
since, in the end, we will renormalize and take the limit $b \to 0$.
Thus we may just collect the leading terms when $T >> b$. Then
\begin{equation}\label{eq:ft2}
{\mathcal F}_t[x]\;\sim\; \frac{e^2}{4 \pi N} \; 
\left[ 
T \ln( \frac{32}{\pi} \frac{1}{e^2 b} ) \;-\; \frac{8}{e^2} 
\ln (\frac{\pi e^2 T}{16}) \right]
\end{equation}
where terms that tend to zero when $b/T \to 0$ have been neglected.
It is then possible to insert (\ref{eq:ft2}) and (\ref{eq:g3}) into
(\ref{eq:g2}), and integrate over $T$, to obtain:
$$
G(x-y) \;=\;\kappa \; (\frac{e^2}{8})^{\frac{2}{\pi N}} \,
\exp\left\{ - \frac{4}{\pi^2 N} \, (\xi -1) f_l(r) \right\}
$$
\begin{equation}
\times \, \int \frac{d^3 p}{(2\pi)^3}\; e^{i p \cdot (x-y) } \;
 \;
( i \not \! p + \Sigma )^{-(1 + \frac{2}{\pi N})}\;, 
\end{equation}
where
\begin{equation}\label{eq:defsigma}
\Sigma \;=\; \frac{e^2}{4 \pi N} \;
\ln\left( \frac{32}{\pi} \frac{1}{e^2 b} \right)\;.
\end{equation}
and $\kappa$ is a numerical constant.

It should be evident from the above that $\Sigma$, modulo a
renormalization that we shall explain in what follows, signals the
existence of a mass for the fermionic field. It may be argued that a
fermion mass appears just because the ribbon regularization breaks
chiral symmetry. However, this only explains the existence of a
perturbative mass, and in fact can (and will) be taken into account by
a careful choice of the finite counterterms. The situation is
analogous to the one in $QED_4$ with a non gauge-invariant regulator:
in general, a mass counterterm is usually required in order to have a
gauge invariant renormalized theory.

On the other hand, a non-zero (renormalized) mass which is
non-perturbative in the coupling constant shall be an {\em a
  posteriori\/} justification for the use of this IR approach.

$\Sigma$ determines the position of the singularity in momentum space,
and it is evident from (\ref{eq:defsigma}) that it needs a
renormalization, since it diverges when $b \to 0$. The renormalized
mass, which we shall denote by $m$, will differ from $\Sigma$ by a
subtraction corresponding to a mass counterterm:
\begin{equation}
m \;=\; \Sigma - \Sigma_0
\end{equation}
where $\Sigma_0$ is the counterterm. Now such a divergence in this
superrenormalizable theory does only occur for the self-energy diagram
which is of order $e^2/N$.  Noting on the other hand that $\Sigma$ may
be rewritten as
\begin{equation}\label{eq:sigma1}
\Sigma \;=\; 
\frac{e^2}{4 \pi N} \; \ln(\frac{1}{b \mu}) \;+\;
\frac{e^2}{4 \pi N} \; \ln(\frac{32}{\pi} \frac{\mu}{e^2})
\end{equation}
where $\mu$ is a mass scale, we see that the divergent part
\mbox{$\frac{e^2}{4 \pi N}\ln(\frac{1}{b \mu})$} is indeed of order
$\frac{e^2}{N}$, as expected in this superrenormalizable theory. Then
we fix the divergent part of the mass counterterm to be such that
\begin{equation}\label{eq:ct}
\Sigma_0^{div} \;=\; \frac{e^2}{4 \pi N} \; \ln(\frac{1}{b\mu})
\end{equation}
so that
\begin{equation}\label{eq:mr}
m \;=\; m_0 \,+\, \frac{e^2}{4 \pi N} \;\ln(\frac{32}{\pi} \frac{\mu}{e^2})\;.
\end{equation}
where $m_0$ is a {\em finite\/} counterterm, which has to be fixed by
a normalization condition. It should of course be of the form:
\begin{equation}
m_0 \;=\; \frac{e^2}{4 \pi N} \, C \;,
\end{equation}
where $C$ is a number of order $1$. We do not want to introduce an
extra mass, scale into the theory, so we have to use $\mu = m + \delta
m$ as the renormalization scale:
\begin{equation}\label{eq:mr1}
m \;=\; m_0 \,+\, \frac{e^2}{4 \pi N} \;
\ln(\frac{32}{\pi} \frac{m + \delta m}{e^2})\;.
\end{equation}
The reason why $\mu$ cannot be just equal to $m$ is the following:
$m=0$ has to be one of the possible solutions to equation
(\ref{eq:mr1}), so that chiral symmetry is not explicitly broken. But
setting $m=0$ in (\ref{eq:mr1}) certainly requires a non vanishing
$\delta m$:
\begin{equation}\label{eq:mr2}
\delta m \;=\frac{\pi e^2}{32} \, \exp (- 4 \pi N \frac{m_0}{e^2} ) \;,
\end{equation}
what determines the relation between $\delta m$ and $m_0$. This is in
fact the renormalization condition that guarantees that the chiral
symmetry is not explicitly broken by the renormalization procedure.
Using relation (\ref{eq:mr2}), we may finally write the expression for
$m$ as a transcendental equation:
\begin{equation}\label{eq:mr3}
m \;=\; \frac{\pi e^2 }{32} \, e^{-C} \; 
\left[ \exp (4 \pi N \frac{m}{e^2}) \,-\, 1 \right]\;.
\end{equation}
where $C$ cannot be determined within the approximation made in
obtaining (\ref{eq:ft1}). Indeed, in this approximation the induced
mass is a constant rather than a momentum dependent function.
Therefore it impossible to decide which is the value of the constant
$C$, related to the dynamical cutoff ($\sim \frac{e^2}{8 N}$), which
is observed in the Schwinger-Dyson equations
approach~\cite{Appelquist:1988sr}.

However, from a simple analysis of (\ref{eq:mr3}), we see that besides
the trivial solution $m=0$, there will be a non-vanishing mass as long
as $N$ remains smaller than a critical value $N_c$, given by:
\begin{equation}\label{eq:mr4}
N_c \;=\; \frac{8}{\pi^2} \, e^C \;.
\end{equation}
This result is consistent with the ones obtained in more standard
settings; for example, the value $N_c = \frac{32}{\pi^2}$
of~\cite{Appelquist:1988sr}, corresponds to $C = \ln 4 \sim 1.39$.  It
is interesting to compare this result with the one one would have
obtained by ignoring the one-loop correction to the photon propagator.
As already mentioned above, there is no induced mass term in such an
approximation, since one finds no term linear in $T$ in
(\ref{eq:ft2}).

We conclude by stressing that this IR approximation scheme looks
promising as a tool to study the chiral phase transition in $QED_3$
like models, in a gauge invariant setting.  However, in order to
obtain the full momentum dependence of the fermion mass, a higher
order approximation than the one used here may be required. Indeed, it
is plausible that fermion recoil effects are necessary to sense that
momentum dependence. This effect can not be appreciated in the
approximation used here since the motion of the fermions is not
perturbed by their interaction with photons \cite{Karanikas:1995zi}.

\section*{Acknowledgements}
We thank Prof.\ Dmitri Khveshchenko por pinpointing a typographical 
error in Eq. (37), in a previous version of this manuscript.
The authors acknowledge the kind hospitality of the members of the
Department of Physics of the University of Oxford, and the support of
CONICET (Argentina).  C.\ D.\ F.\ was supported by a Fundaci\'on
Antorchas grant.


\end{document}